\documentclass[preprint,showpacs,preprintnumbers,amsmath,amssymb]{revtex4}

\usepackage{graphicx}% Include figure files
\usepackage{dcolumn}% Align table columns on decimal point
\usepackage{bm}% bold math
\usepackage{natbib}

\begin{document}

%\preprint{APS/123-QED}

\title{Ordered and disordered dynamics in monolayers of rolling particles }% Force line breaks with \\

\author{Byungsoo Kim}
 \email{kimb@math.colostate.edu}
 \author{Vakhtang Putkaradze}
 \affiliation{Department of Mathematics, Colorado State University, Fort Collins CO 80235, USA}%Lines break automatically or can be forced with \\

\date{\today}% It is always \today, today,

\begin{abstract}
We consider the ordered and disordered dynamics for monolayers of rolling self-interacting particles with an off-set center of mass and a non-isotropic inertia tensor.  The rolling constraint is considered as a simplified  model of a very strong, but rapidly decaying bond with the surface, preventing application of the standard tools of statistical mechanics.  We show the existence and nonlinear stability of ordered lattice states, as well as disturbance propagation through and chaotic vibrations of these states. We also investigate the dynamics of disordered gas states and show that  there is a surprising and robust linear connection between distributions of angular and linear velocity for both lattice and gas states, allowing to define the concept of temperature.
\end{abstract}
%%% Todo
\newcommand{\bs}{\mathbf{s}}
\newcommand{\bp}{\mathbf{p}}
\newcommand{\bE}{\mathbf{E}}
\newcommand{\bV}{\mathbf{V}}
\newcommand{\bF}{\mathbf{F}}
\newcommand{\bT}{\mathbf{T}}
\newcommand{\bGamma}{\boldsymbol{\Gamma}}
\newcommand{\bOmega}{\boldsymbol{\Omega}}
\newcommand{\rem}[1]{}
\newcommand{\todo}[1]{\vspace{5 mm}\par \noindent
\framebox{\begin{minipage}[c]{0.45 \textwidth}
\tt #1 \end{minipage}}\vspace{5 mm}\par}

\pacs{45.50-j,02.20.Ns}% PACS, the Physics and Astronomy
\maketitle

%\section{Introduction}
%%%%%%%%%%%%%%%%%%%%%%%%%%%%%%%%%%%
\noindent
{\bf Introduction}
Molecular monolayers are
playing an ever increasing role in technology as they allow manipulation of contact
properties of materials in a precise and controlled manner.
There have been considerable effort in analyzing the structure of
molecular monolayers; in particular, self-assembled polymeric
monolayers \cite{Ul1996,Ma2002} and water monolayers with TIP5P model \cite{Za2003}.
Previous theoretical work on molecular monolayers has concentrated
on either static studies of monolayer structures
\cite{Ca2009,Ul1996} or molecular dynamics simulations where each
molecule is moving under the forces and torques from surrounding
molecules and the substrate \cite{Ma2002,Ku2010,Za2003,Ku2006}. The direct
molecular simulations employed in these papers were quite successful
in explaining various properties of liquid water, \emph{e.g}
formation of contact angle \cite{Ma2002} or anomalous properties for nano-confined water \cite{Ku2006,Ku2010}. In the latter papers, the confinement of
water molecules to a very thin layer was achieved by physically
constraining the bulk of water to a very narrow (nm) layer by the
rigid surface on both sides. On the other hand, experimental data
demonstrate the presence of water monolayer on a Si surface under
normal conditions, which is due to the strong bond between the water
molecules and the substrate \cite{Mi1998,Ca2009}. Such a bond is
difficult to account for in the atomistic molecular simulations of
TIP5P type. On the other hand, one would like to keep the relative
simplicity of atomistic models without introducing extra coupling
bonds, the nature of which is not well understood.

\noindent
The purpose of this Letter is to suggest a simple yet physical way of introducing a strong coupling bond between the monoloayer molecules and the substrate as a rolling constrant. We assume that the bonds between the substrate and the molecule are very short ranged and thus act on the part of the molecule in close contact with the substrate atoms. Physically, that leads to the fact that while the molecule itself is moving, its point of contact with the substrate is stationary.
This invokes the analogy with a classical problem of rolling body on the surface with perfect friction, appearing when the bond  molecule-substrate is infinitely strong at contact point, but decays rapidly away from substrate. In reality, if that bond is large but finite, the motion will be a combination of sliding and rotating. However, the theory of sliding and rolling is not yet well developed \cite{Re1994,Bl2005}, and so in this paper we will restrict ourselves to the simplest possible realization when the friction is perfect. Thus, we consider the dynamics of a monolayer consisting of rolling molecules that are self-interacting by the long-range interactions (Lennard-Jones and electrostatic), while the influence of the boundary is limited to restricting the motion to the perfect rolling dynamics. While the rolling dynamics may seem too idealized, it has actually been observed in the context of functional nano-structures \cite{Shi2006}.  More complex models of molecule-substrate interaction are possible, resulting in the gravity-like component in the force acting on the ball, but such interactions will not be considered here.

The study of rolling motion of rigid bodies has a  long history in the context of classical mechanics \cite{Bl2003,Bl2005,Go2001}. However, the study of collective motion of rolling particles has not been undertaken. The non-holonomic rolling constraint  is a major obstacle in the way of constructing statistical mechanics for the rolling particle systems  \cite{Ta2005,Go1979,Ku1999,La1980}. This Letter is devoted to defining ordered and disordered states of the rolling particle systems (akin to solid and gas/liquid), and computing statistical physics  concepts for such systems. We also show the way rolling affects propagation of phonons through a rolling particle lattice, which opens the way to experimentally confirm the molecular rolling motion.
%\section{Simulation setting}
%\subsection{Dynamics of rolling balls}

\begin{figure}[h]
  \begin{center}
    \scalebox{.20}{\includegraphics{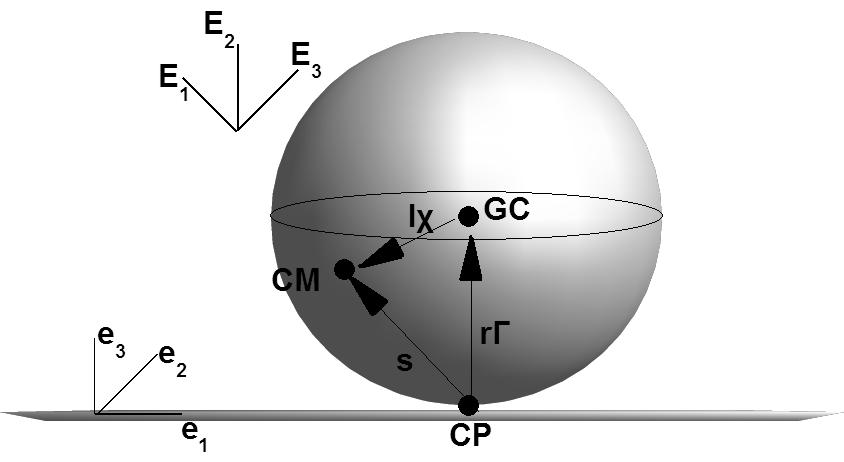}}
    \end{center}
    \vspace{-20pt}
    \caption{Schematic and definitions of an offset rolling ball dynamics. }
\label{cball}
\end{figure}

\noindent
{\bf Setup of the dynamics}
The rolling particles are simulated as identical spherical rigid bodies of radius $r$ all having the same mass $m$ and the moments of inertia tensors. The center of mass is assumed to be at a position different from the geometric center, as illustrated on Fig.~\ref{cball}. The notation used in this paper is as follows: $\bGamma_i$ is the unit vector pointing to the geometric center (GC), $\bOmega$ is the angular velocity \emph{in the ball's coordinate frame} and $\bs_i$ is the vector pointing to the center of mass (CM).
The equations of motion for an individual ball come from the well-known Chaplygin's equations \cite{Bl2005,Bo2002}.  Under cetrain symmetry conditions, Chaplygin's ball is completely integrable has three integrals of motion: one is total energy that is the easily understandable physical quantity. The other two constants of motion are Routh and Jellet integrals that are harder to explain in terms of elementary physics \cite{Gr2000}. The equations of motion for the $i$-th ball are as follows:
\begin{align}
 \left( \frac{d}{dt} +\bOmega^i \times \right) &
 \left(I^i \bOmega^i + m \bs^i \times (\bOmega^i \times \bs^i) \right)
 \nonumber
 \\
 &-m^i \dot{\bs}^i\times(\bOmega^i \times \bs^i)  = s^i \times \bF^i +\bT^i \label{motion1} \\
 \frac{d \bGamma^i}{dt} = - \bOmega^i & \times \bGamma^i  \, .  \label{motion2}
\end{align}
 Here, $I^i$ are the eigenvalues of the tensor of inertia and  $\bF^i$ and $\bT^i$ are the total force and torque acting on the $i$th particle. These forces and torques include the inter-particle interactions.
 The rolling  constraint for velocity of the center of mass $\bV^i$, written as
$
\bV^i=\bOmega^i \times \bs^i
$
cannot be reduced to an equation between configuration variables only, and is thus non-holonomic \cite{Bl2003}.  The Hamiltonian approach for the ensemble of rolling ball cannot be defined, and  thus it is impossible to construct  a meaningful approach for this system by simply extending the classical statistical physics to this case. Note that neither linear nor angular momenta are conserved, neither for individual particles nor for the whole system.  The total energy of the system, however,  is conserved.

The inter-molecular interactions causes the interchange of linear and angular momentum as well as  energy between particles. In the simulation of rolling water molecules, the Lennard-Jones (LJ) potential $V_{LJ}$ and the dipole potential $V_d$ are considered. The dipole $\bp_i$ is positioned at the center of mass on each spherical particle. The  interactions we consider are defined as follows:
\begin{eqnarray}
V_{LJ} &=2 \epsilon &  \sum_{i \neq j}  \left( \frac{\sigma^{12}}{r_{ij}^{12}}-\frac{\sigma^{6}}{r_{ij}^{6}} \right)  \label{VLJ} \,, \\
V_{d} &=& \frac{1}{8\pi \epsilon}\sum_{i \neq j}  \bp_i \cdot  \left( \frac{3 (\bp_j \cdot \hat{r_{ij}}) \hat{r_{ij}}-\bp_j}{r_{ij}^3} \right) \, .  \label{Vd}
\end{eqnarray}

%\noindent
%{\bf Simulation parameters}
To be concrete, in the computations of collective dynamics for rolling particles, we use the parameters and interactions relevant to  water molecule monolayers  by choosing the mass $m=2.991 \cdot 10^{-23}$ g, moments of inertia $(I_1,I_2,I_3)=(0.2076, 0.1108, 0.3184) \cdot 10^{-39}$ g$\cdot$cm$^2$, radius $r=1$ \AA, displacement of center of mass from the geometric center $\ell=0.068$\AA, dipole moment $6.17 \cdot 10^{-30}$ (C $\cdot$ m), LJ radius $\sigma=3.165$ \AA  \, and energy $\epsilon=0.650$kJ/mol. These values correspond to the parameters of a water molecule  \cite{Be1981}. For convenience, we choose the angular velocity scale $\tilde \omega=10^{13}$rad/s. The rolling constraint then introduces the scaling of velocity to be $\tilde v =r \tilde \omega=10$cm/s.   Given a different set of parameters, the details of our computations will be different, but the methods and results outlined here hold for other molecules as well. For the set of parameters considered here, neither Routh nor Jellet integrals for each ball are conserved.
%The displacement of the center of mass from the geometric center  is assumed as the displacement from the center of oxygen atom.

%\subsection{Simulation scenario}
\noindent
{\bf Stationary states: a crystalline lattice}
The existence of stationary states for the system of rolling particles depends on the presence and orientation of the dipole moment. Suppose for now that the dipole moment is absent, and the only interaction between the molecules is Lennard-Jones. Suppose also for the moment that the particles are not moving, and are aligned so both geometric center and center of mass are along $\bGamma$, the unit vector pointing upward from the contact point.  It is easy to see then that there is an equilibrium configuration so that the centers of mass are arranged at a distance close to $\sigma$, the equilibrium distance of LJ potential. For the case of two and three particles, there is an equilibrium configuration where the particles are arranged at exactly the distance $\sigma$ from each other.

\begin{figure}[h]
  \begin{center}
    \scalebox{.25}{\includegraphics{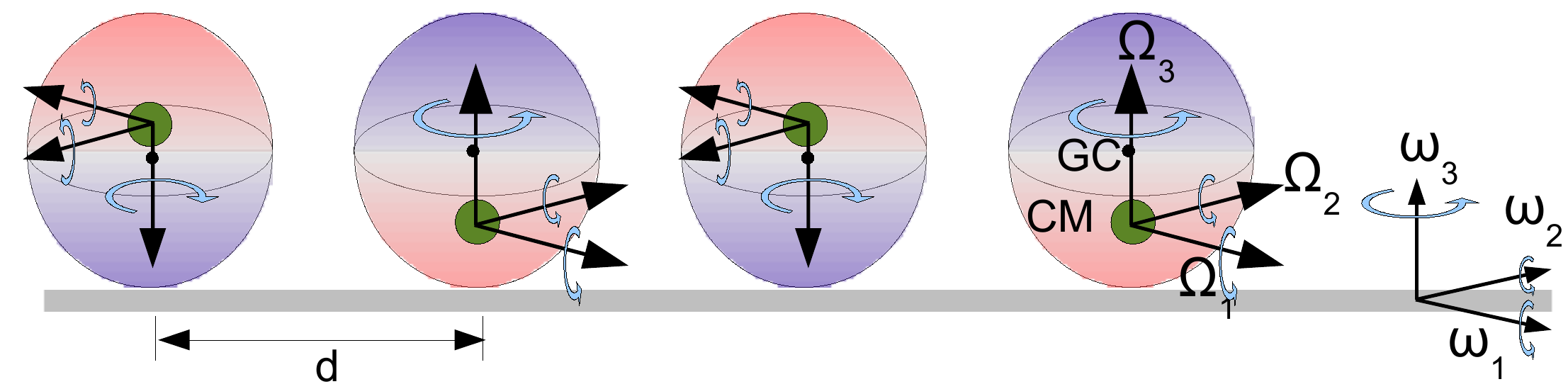}}
    \end{center}
        \vspace{-20pt}
    \caption{Schematic representation of stationary states.}
\label{fig:statstates}
\end{figure}

Let us suppose now that these particles are spun with angular velocity $\bOmega_i=(0,0,\Omega_i)$ that is pointing upwards. In the absence of nonlocal interaction this will be a neutrally stable state for each particle, although friction forces may destabilize such equilibria, similar to familiar phenomenon of the tippe top \cite{Na2008}. Because LJ forces are exactly at balance, independent of the values of $\Omega_i$, this will be an equilibrium state. In the presence of the dipole moment, finite lattices cease to be equilibrium configurations. However, if the dipole moment points exactly along the line from the geometric center to the center of mass (which is the case for the particles considered here), an infinite regular lattice, possessing high symmetry (like a triangular or square lattice) will still be an equilibrium state. The dipole moments can either be arranged in the same directions or be alternating. However, the states with the dipole moments all pointing in one direction are unstable both linearly and nonlinearly. Thus, in what follows we concentrate on the states with the alternating dipole moments as shown on Figure~\ref{fig:statstates}.
\vspace{-10pt}

\begin{figure}[h]
  \begin{center}
      \scalebox{.25}{\includegraphics{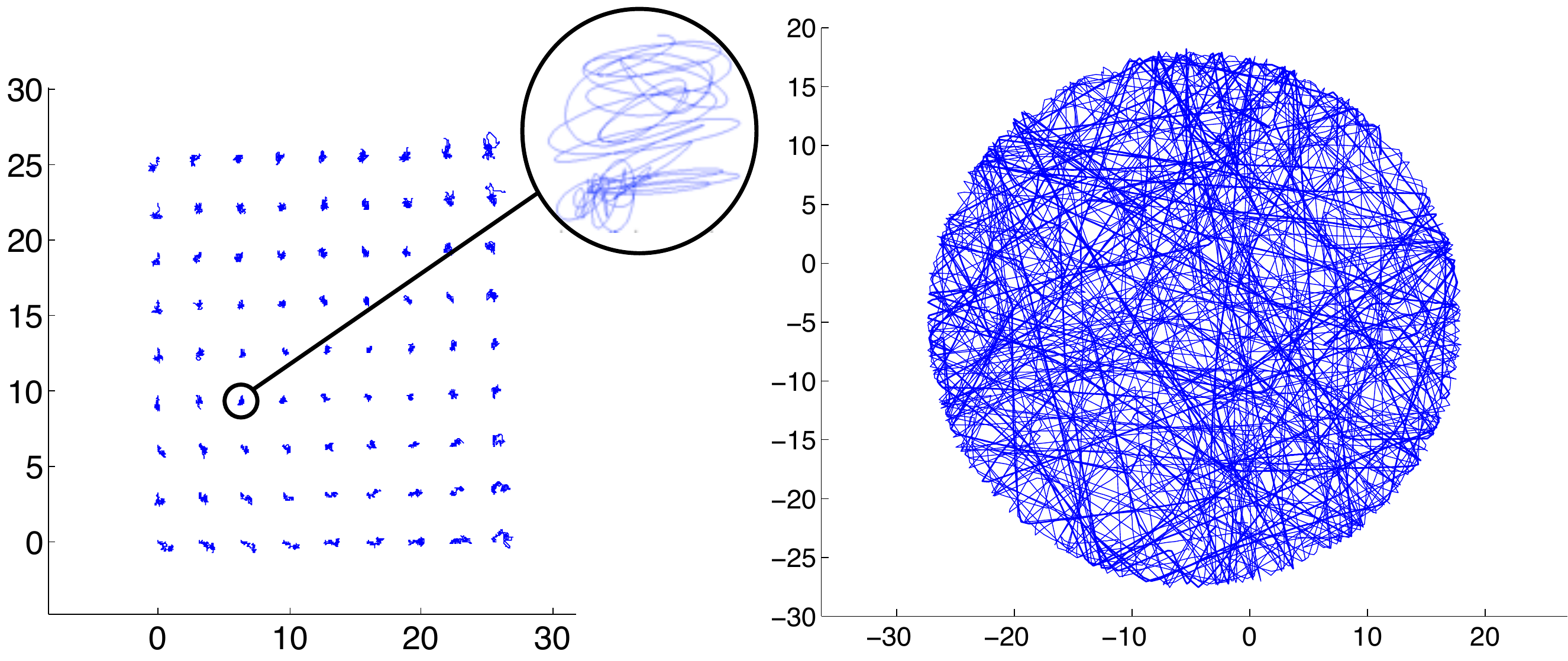}}
    \end{center}
        \vspace{-20pt}
    \caption{Left: trajectory of the center of rolling particles in lattice model, with the blow-up showing the trajectory of an individual particle. The finite lattice is not an exact stationary state, so the vibrations are most apparent at the edges.  Right: trajectories of a gas state in a circular container. }
\label{lattraj}
\end{figure}

The question about the stability of these states immediately arises. While a detailed investigation of the stability is a complex issue beyond the scope of this Letter, a simple physical explanation can be made showing that these states are linearly unstable. Suppose for simplicity that all rotation rates are the same, $\bOmega_i=\bOmega_0$. A small perturbation in the spinning stationary state will result to a precession of each particle, which according to the rolling constraint will happen at the same rate as the rotation frequency. The distance between particles will also change with the same frequency, and through  the LJ interaction (\ref{VLJ}) both particles will experience parametric resonance. Similar argument applies to different rotation rates, and more particles in a lattice. Thus, all lattice states are linearly unstable. However, these states are \emph{nonlinearly stable} at least for some of the configurations and initial conditions we have investigated, corresponding to low initial energies.  In Fig.~\ref{lattraj}, left, we show the positions of the centers of 81 rolling balls situated in a rectangular lattice, over a long simulation time. While each particle remains close to its equilibrium position, the individual trajectories (blow-up on the same Figure) are chaotic and are strongly reminiscent of thermal vibrations in lattices.
\vspace{-10pt}

\noindent
{\bf Lattice dispersion relation}
In order to suggest a possible experimental verification of rolling motion, we suggest measuring the propagation of disturbances through the lattice. The derivation of dispersion relation for a lattice of spinning particles in general case is difficult, as it involves linearization about oscillating base states. Here, we present the analysis of the rolling and rocking lattice with all the particles being stationary in the base state, \emph{i.e.} $\Omega_i =0$.  Assuming an infinite square lattice as described above, we arrive to the following dispersion relation for the propagation of disturbances of the form
$e^{ -i \omega t+ i k_x x +i k_y y}$,  $(k_x,k_y)$ being  the wave vector:
\begin{eqnarray}
\left\{ \frac{m}{K} \big(1+\zeta_1)\omega^2-4+2\cos(k_x a)(1+ \cos(k_y a)) \right\} \times & \nonumber \\
\left\{ \frac{m}{K} \big(1+\zeta_2)\omega^2-4+2\cos(k_y a)(1+ \cos(k_x a)) \right\} & \label{disp}\\
- 4 \sin ^2(k_x a) \sin^2 (k_y a) =0 \, ,  \nonumber &
\end{eqnarray}
where $\zeta_i=I_i/\big(m (r+ \ell)^2 \big)$,  $K=d^2 V_{LJ}/dr^2$ is the spring constant of the $LJ$ potential and $a$ is the periodicity of the square lattice .  Note that (\ref{disp}) differs from the standard dispersion relation for a square lattice of springs only by the dimensionless coefficients $\zeta_i$, incorporating the effects of rolling. For our values of parameters,
$\zeta_i \simeq 0.1 - 0.2$. Thus, the rolling constraint affects the speed of sound by about 10-20\%, which should be a measurable difference.

\rem{ %%%%BEGIN REM
This  dispersion relation (\ref{disp}) only makes sense within the periodicity domain $0 \leq (k_x,k_y) \leq 2 \pi/a$. An example of two roots of $\omega(k)$ is given in Fig.~\ref{fig:dispersion}. The frequency $\omega$ is always real.
\begin{figure}[h]
  \begin{center}
    \scalebox{.30}{\includegraphics{dispersion_omega21-small.png}}
     \scalebox{.30}{\includegraphics{dispersion_omega22-small.png}}
    \end{center}
     \vspace{-20pt}

    \caption{Dispersion relations given by two different roots of (\ref{disp}). }
\label{fig:dispersion}
\end{figure}
\vspace{-10pt}

The dispersion relation for a lattice consisting of non-rolling balls connected by LJ springs can be easily obtained from (\ref{disp}) by setting $I_1=I_2=0$, $l=0$. For the values of parameters chosen here, $\omega(k)$ for rolling and non-rolling balls differ by about 20\%, which should be a noticeable and measurable difference.
} %%%END REM

\noindent
{\bf Disordered states: statistical analysis}
For large initial energies, the lattices become unstable, and in the absence of external boundaries the particles scatter to infinity, making the concepts of statistical physics meaningless. Thus, for large energies leading to gaseous states, we perform the simulations in a round potential well with sharp walls, forcing the particles to remain within a circle. An example of such simulation with 16 particles is shown in Figure~\ref{lattraj}, right.

The first step towards considering this system as a statistical physics model is to investigate the distribution of linear and angular velocities. For an ideal gas in 3D, the Maxwell-Boltzmann distribution for velocities is
\begin{equation}
f_v(\mathbf{v}) = \left( \frac{m}{2\pi kT}\right)^{3/2} \exp \left(  - \frac{m \mathbf{v}^2}{2kT}\right) \,,
\label{MBdist}
%f_v(v_i) &=& \sqrt{\frac{m}{2\pi kT}} \exp \left( - \frac{mv_i^2}{2kT}\right) \,, \\
%f_E &=& 2 \sqrt{\frac{E}{\pi (kT)^3}} \exp \left(-\frac{E}{kT}\right) \,.
\end{equation}
and similar for the rotational degrees of freedom. The implicit assumption in (\ref{MBdist}) is that the distribution of velocities in each direction is normal with \emph{the same} width that defines the temperature $T$ of the system.   However, in the system of rolling particles there is no reason for (\ref{MBdist}) to work: first, the rotational and translational components are coupled through the rolling constraint and second,  even though the rolling motion is along the plane, each ball undergoes three-dimensional motion.

\begin{figure}[h]
  \begin{center}
    \scalebox{.26}{\includegraphics{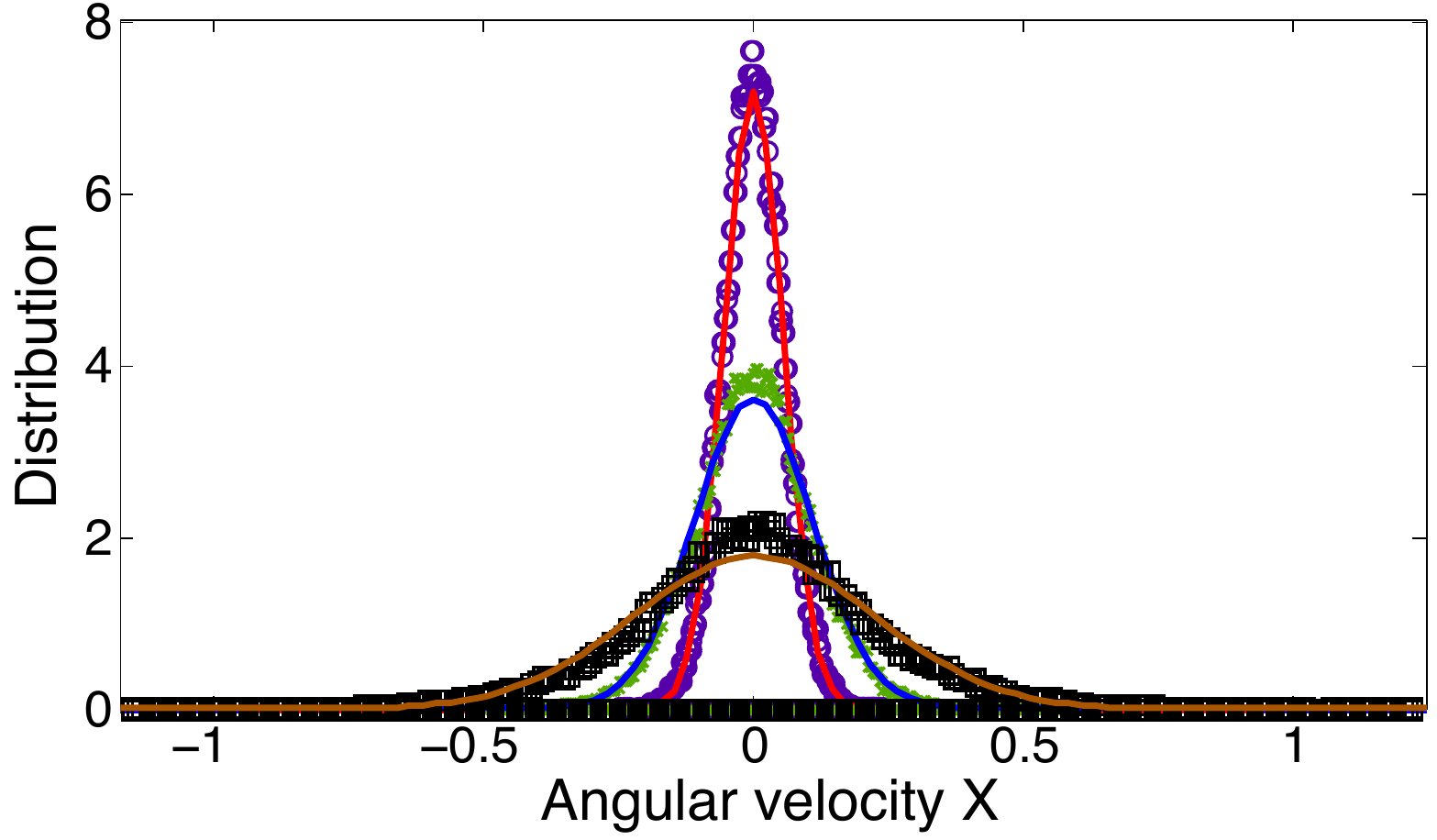}}
    \scalebox{.26}{\includegraphics{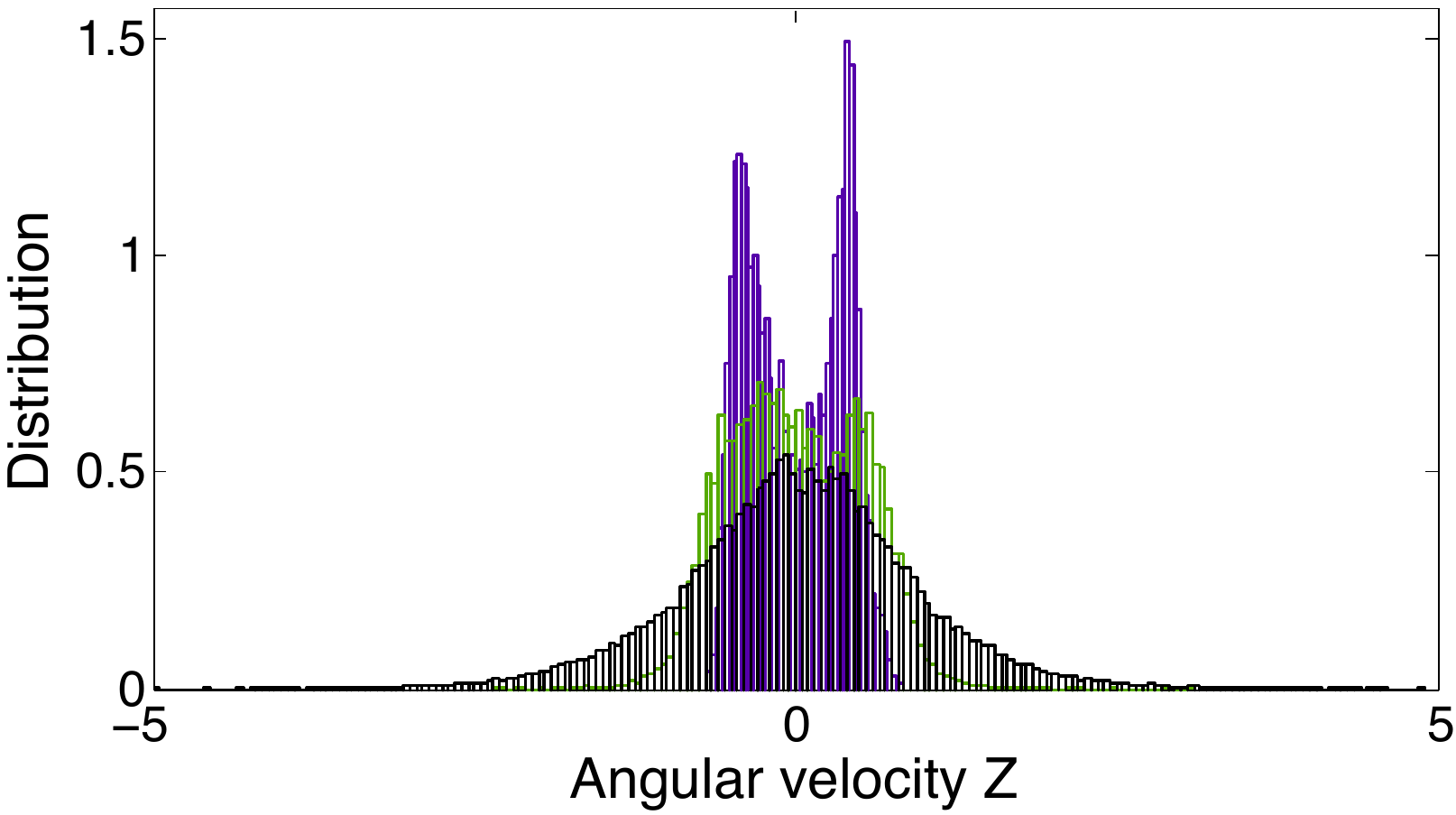}}
    \end{center}
    \vspace{-20pt}
    \caption{Distributions of $\omega_x$ (left) and $\omega_z$ (right) for different energies of the system. Solid curves represent fits to normal distributions.}
\label{fig:omega3}
\end{figure}

Figure~\ref{fig:omega3} shows several examples of the distribution for several values of total energy of the system.
On the left side of this Figure, we plot the distribution of $\omega_x$, which is identical to the distribution of $\omega_y$.
These distributions are always very close to  normal, which are shown with solid curves.
The right side of this Figure shows distributions of $\omega_z$ for the same values of the energy, and it is apparent that this distribution is not normal. The distributions of $v_x$, $v_y$ and $v_z$ show the same tendencies.  It is important to note, however, that the variances (in proper units) of normally distributed angular and linear velocities are \emph{not} the same, and thus there is no straightforward definition of temperature for this system.
\vspace{-10pt}
We shall also note that due to the non-normal nature of the distribution in one of the components and the non-holonomic coupling between angular and linear velocities, the kinetic energy distribution does not follow Maxwell-Boltzmann law.
In addition, the kinetic energies of rotational and translational motion are not equal (even in a statistical sense), so the equipartition of energy in our system does not hold.

\begin{figure}[h]
  \begin{center}
    \scalebox{.18}{\includegraphics{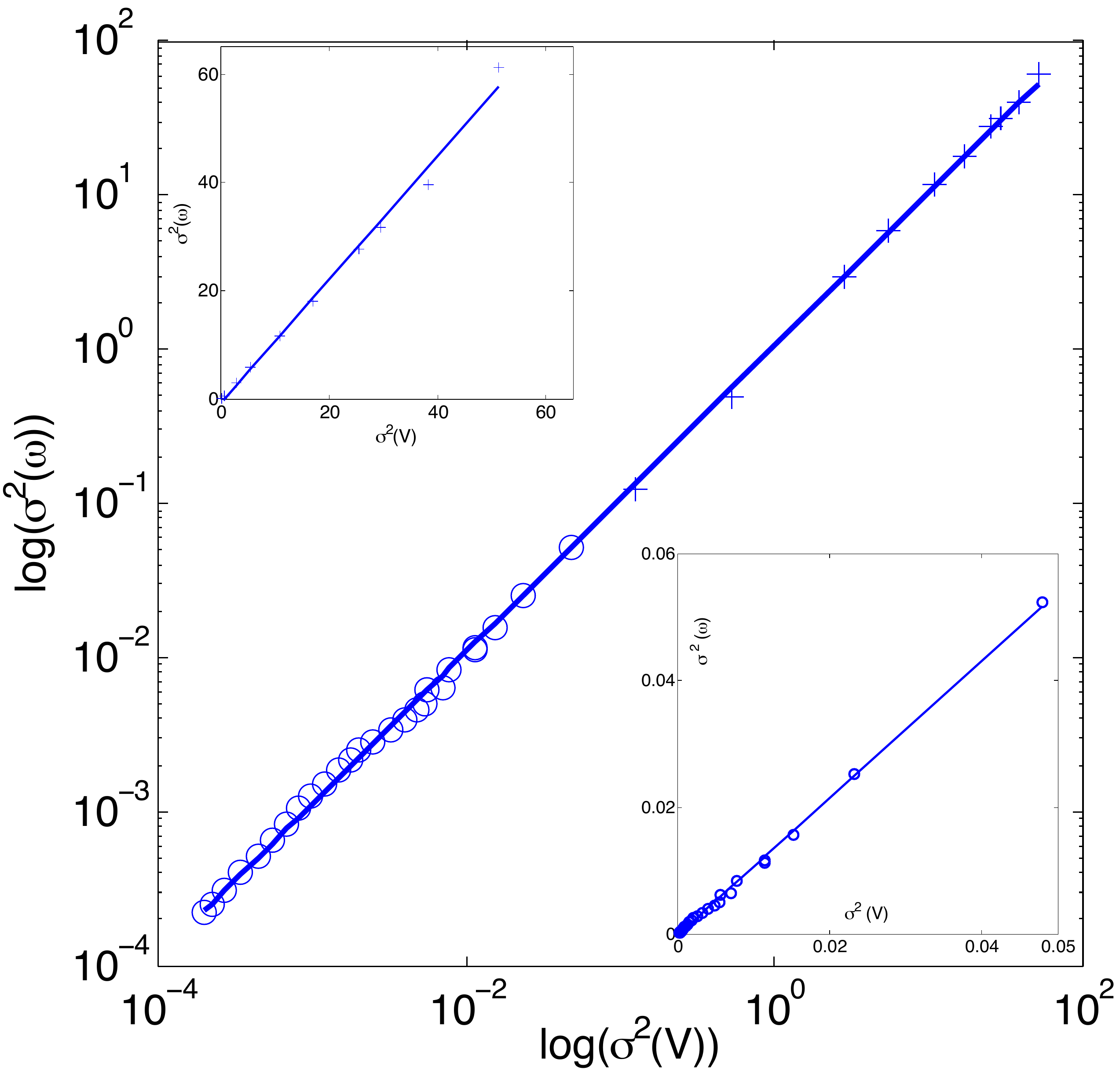}}
    \end{center}
\vspace{-20pt}
    \caption{
    Linear relationship between variances $\sigma_v$ and $\sigma_\omega$ for lattice states (circles) and gas states (lines). Common linear fit to both sets of data is also shown.  The linear plots for each regime are shown as inserts.  }
\label{fig:scaling}
\end{figure}

\noindent
{\bf Temperature as scaled variance}
Clearly, the rolling constraint prevents a straightforward statistical physics description of the rolling particle systems. Nevertheless, there is a surprising relation  that connects the variances $\sigma$ obtained from the angular ($\sigma_\omega$) and linear ($\sigma_v$) velocities. Namely, we observed that for all values of in our numerical experiments,  there is a surprising well-behaved linear relationship
$\sigma_\omega= k \sigma_v$, with the coefficient $k \simeq 1.08$ depending only on the parameters on the system (geometry of the ball, center of mass position \emph{etc}) but not on anything else. That relationship is valid for both lattice and gas states, as  shown in Figure~\ref{fig:scaling}. We note that there is no \emph{a priori} reason for such relationship to exist, but the surprising robustness of this law leads us to believe that it could be taken as one of the postulates in future development of statistical mechanics for non-holonomic gas.
  \vspace{-10pt}

\begin{figure}[h]
  \begin{center}
     \scalebox{.115}{\includegraphics{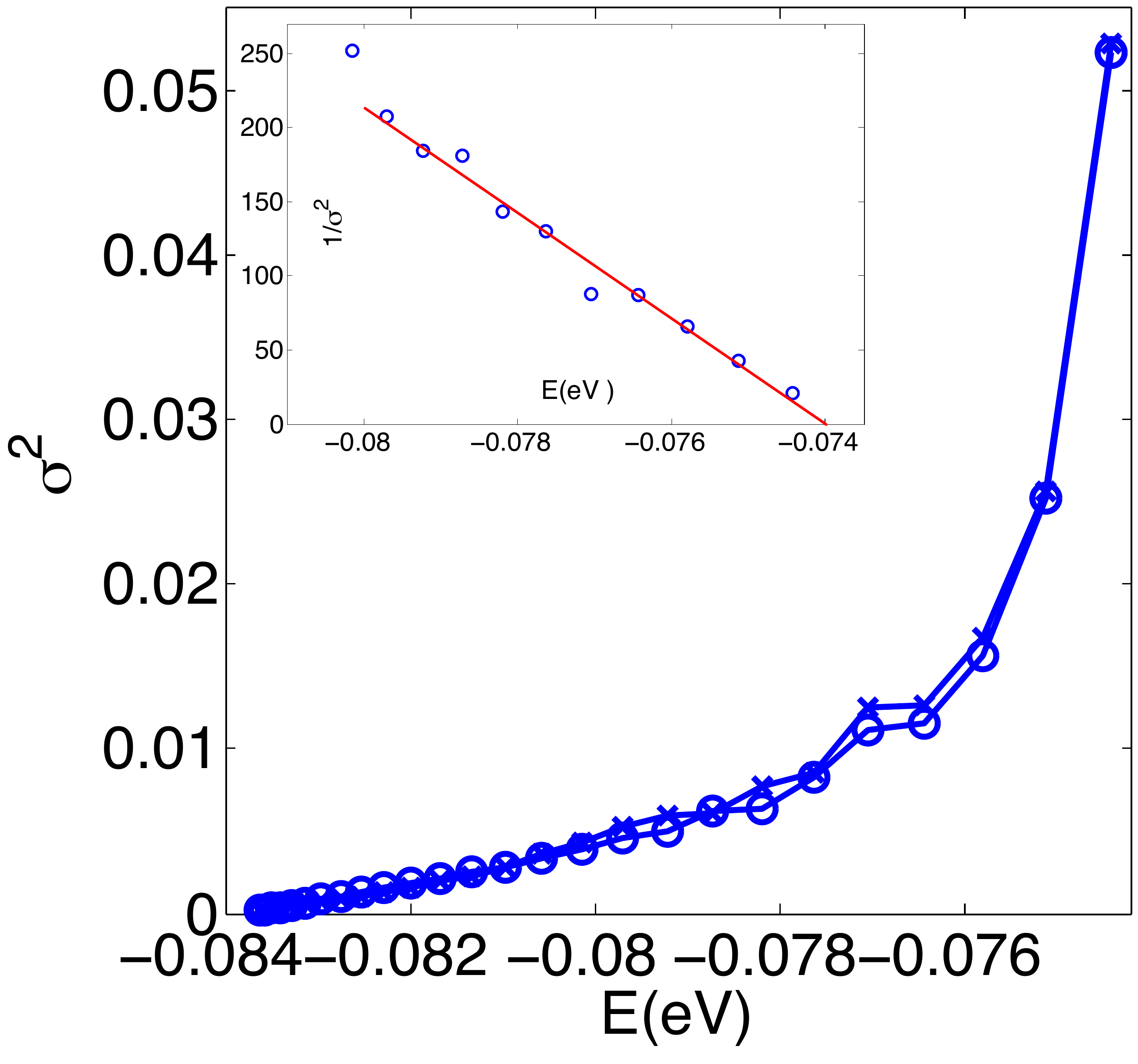}}
     \scalebox{.23}{\includegraphics{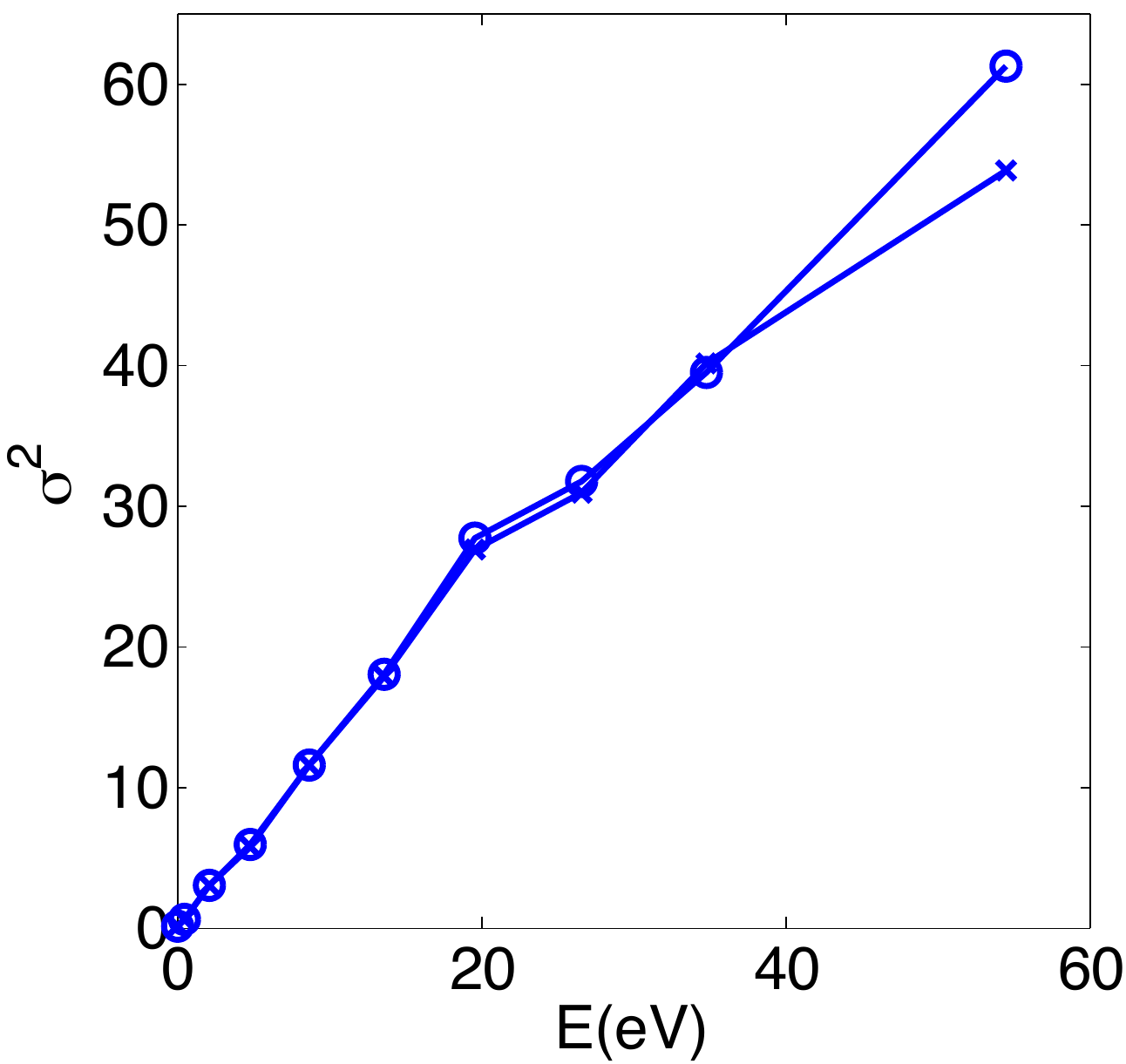}}
    \end{center}
\vspace{-20pt}
    \caption{Variances vs energy in the lattice state (left) and gas state (right). Circles:  $\sigma^2(\omega)$-distribution; crosses: $\sigma^2(v)$-distribution multiplied by $k=1.08$. Left insert: $1/\sigma^2$ vs $E$.  }
\label{fig:energy}
\end{figure}

Using this surprising relation we can define the temperature as the variance of either horizontal linear or angular velocity, or any linear combination of those. Since the temperature is defined up to a constant, all these definitions lead to the same results up to a scaling factor. Thus, on Figure~\ref{fig:energy},  we plot the scaled variances vs total energy of the system. If we think of the variance $W$ as being proportional to temperature, we observe that the total energy $E$ is directly proportional to the temperature,which is reminiscent of ideal gas. Thus, in spite of complexities of the rolling systems, some aspects from the ideal gas remain. We shall note that since in simulations we use real balls interacting with LJ potential, a more detailed simulation should show the effects of the finite size of the particles, but our simulations do not allow a  reliable investigation of these effects.

Using the concept of temperature as scaled invariance, we can now define thermodynamically meaningful equations of state. On Figure~\ref{fig:energy}, we plot the temperature as a function of the energy per particle. The equation of state for the lattice state (left panel of the Figure) ceases to exist for small negative energies as the lattice becomes unstable and ceases to exist. The exact nature of the destruction of lattice state is as yet unclear, but the divergence of the variance $\sigma^2 \sim (E_0-E)^{-1}$ (shown in the insert) indicates the presence of a phase transition.

 \vspace{-10pt}

\noindent
{\bf Acknowledgements} The authors were partially supported by grants NSF-DMS-0908755 and  HDTRA1-10-1-0070. We also benefitted from fruitful discussions with Profs. D. D. Holm, P. Vorobieff  and C. Tronci.

\bibliography{bsoobib}

\end{document}